\documentclass[letterpaper,twocolumn,10pt]{article}
\usepackage{usenix2019_v3}
\usepackage{subcaption}
\usepackage{xcolor}
\usepackage{graphicx}
\usepackage{listings}
\begin{document}

\date{}

\title{\Large \bf The HTTP Garden: Discovering Parsing Vulnerabilities in HTTP/1.1 Implementations by Differential Fuzzing of Request Streams}

\author{
    {\rm Ben\ Kallus} \\
    Dartmouth College \and
    {\rm Prashant Anantharaman} \\
    Narf Industries \and
    {\rm Michael E. Locasto} \\
    Narf Industries \and
    {\rm Sean W. Smith} \\
    Dartmouth College
} 

\maketitle

\begin{abstract}
    HTTP/1.1 parsing discrepancies have been the basis for numerous classes of attacks against web servers (e.g., \cite{watchfire,hot,cpdos}).
    Previous techniques for discovering HTTP parsing discrepancies have focused on blackbox differential testing of HTTP gateway servers (e.g., \cite{treqs,hdiff,frameshifter}),
    despite evidence that the most significant parsing anomalies occur within origin servers~\cite{grenfeldt}.
    While these techniques can detect some vulnerabilities, not all parsing discrepancy-related vulnerabilities are detectable by examining a gateway server's output alone.
    Our system, the HTTP Garden, examines both origin servers' interpretations and gateway servers' transformations of HTTP requests.
    It also includes a coverage-guided differential fuzzer for HTTP/1.1 origin servers paired with an interactive REPL that facilitates the automatic discovery of meaningful HTTP parsing discrepancies and the rapid development of those discrepancies into attack payloads.
    Using our tool, we have discovered and reported over 100 HTTP parsing bugs in popular web servers, of which 68 have been fixed following our reports.
    We designate 39 of these to be exploitable.
    We will release the HTTP Garden to the public on GitHub under a free software license to allow researchers to further explore new parser discrepancy-based attacks against HTTP/1.1 servers.
\end{abstract}

\section{Introduction}

    Despite the increasing adoption of HTTP/2 and HTTP/3, HTTP/1.1 remains the most popular of the HTTP protocols~\cite{cloudflare}.
    This is partly because HTTP/2 and HTTP/3 are optimized for efficient connection reuse, so their performance benefits are less visible for single-request workloads, such as simple API calls~\cite{varvello2016web, zimmermann2017http}.
    Further, while it is common to see communication between browsers and CDNs use these binary protocols, communication between CDNs and their backing servers most commonly uses HTTP/1.1.
    This may very well change in the future, but given that many of the most popular web servers, including Apache httpd and NGINX, still support 1989's HTTP/0.9, it seems unlikely that support for HTTP/1.1 would be going away in the near future.
    Consequently, as long as HTTP/1.1 remains supported by web servers, attacks against it will remain viable.

    HTTP/1.1 comes from a time in which the security consequences of Postel's Robustness Principle were not well-known~\cite{postel1981rfc0793, sassaman2012patch}.
    Although the HTTP/1.1 RFCs define a formal grammar for the protocol, they require only that senders of HTTP messages conform to the grammar.
    In all but a few cases, receivers are free to interpret malformed messages as they see fit.
    Further, the RFCs also designate some behaviors as explicitly implementation-defined.
    As a result, it comes as no surprise that substantial parsing discrepancies exist between nearly all pairs of HTTP/1.1 implementations.

    When servers are chained together, these discrepancies can have consequences.
    For example, many web services route incoming requests through a cache server to avoid regenerating the same page multiple times.
    However, when the cache server and origin server differ in their interpretation of an incoming request's URI, the cache might store data that the origin server didn't expect~\cite{cpdos}.
    In the worst case, such discrepancies can allow attackers to bypass access controls and populate the cache with resources of their choosing.

    Other discrepancies may be entirely benign.
    Many such discrepancies are permitted by the standards, and any differential testing tool must accommodate these.
    For example, some HTTP implementations support implicitly-empty message bodies in \texttt{POST} requests, but many will reject such requests with status code \texttt{411}.
    Both of these behaviors are permitted by the RFCs, and since \texttt{411} is not a cacheable response code, this cannot be used for cache poisoning.

    Still others of these discrepancies are harder to categorize.
    For example, the HTTP/1.1 RFCs require that chunk size lines be delimited with \texttt{CRLF}, but many HTTP/1.1 servers accept chunk size lines delimited by \texttt{LF} alone.\footnote{Note that other types of lines in the HTTP/1.1 protocol \textit{are} permitted to be delimited by \texttt{LF} alone.}
    While these servers are violating the RFCs, this violation is only consequential if one such server is deployed in a chain along with a second server that both differently interprets \texttt{LF} in that position and forwards the message without normalizing such \texttt{LF} bytes into \texttt{CRLF} sequences.
    We are unaware of a server that belongs to this second category, and it is possible that no such server exists.
    However, it is conceivable that such a server may be written in the future, rendering exploitable a bug that was previously benign.

    The state of the art for discovering exploitable discrepancies between HTTP servers is blackbox differential testing. %
    With this technique, inputs are generated and sent to servers, their outputs are compared, and inputs that cause servers to respond differently are reported to the user.
    These discrepancies are rarely exploitable, so significant human effort must be spent combing through fuzzer output and testing its usefulness.

    We present a new system for discovering exploitable HTTP parsing discrepancies: the HTTP Garden.
    The HTTP Garden is a human-in-the-loop differential fuzzer for HTTP.
    Prior work has focused on enumerating discrepancies without automatically evaluating their exploitability.
    We propose two new metrics to differentiate exploitable discrepancies from harmless ones: discrepancy \textbf{meaningfulness}, and discrepancy \textbf{durability}.
    A \textbf{meaningful} discrepancy is one that exists after taking into account the optional requirements of the standards.
    A \textbf{durable} discrepancy is one that survives normalization passes that are typically applied by consumers of the protocol.

The core contributions of our paper are summarized as follows.\footnote{We will add the GitHub URL to the repository in the de-anonymized paper.}
\begin{itemize}
    \item The HTTP Garden is an interface for researchers to study how different proxies and servers interpret HTTP/1.1 messages.
    Our testbench provides the foundation for differential analysis of these servers by presenting side-by-side interpretations of the HTTP/1.1 messages while abstracting away several manual tasks involved in HTTP/1.1 differential analysis (Section~\ref{sec:http-garden}). 
    \item We developed the first-of-its-kind coverage-guided differential fuzzer for HTTP, along with two novel techniques to filter out \textit{unimportant} discrepancies. (Section~\ref{sec:diff-fuzz}).
    \item We leveraged our fuzzer and testbench to discover over 100 bugs in HTTP implementations. Several of these vulnerabilities went undiscovered despite previous attempts at examining HTTP servers and proxies~\cite{hdiff, treqs, frameshifter}. We used our collection of bugs to construct attack payloads that affect numerous common HTTP origin-gateway combinations (Section~\ref{sec:findings}).
\end{itemize}

\section{Background}
    \subsection{HTTP/1.1}
        HTTP/1.1 is a request/response protocol conducted over either TLS or plaintext TCP.
        HTTP/1.1 was first specified in 1997 with RFC 2068~\cite{fielding1997rfc2068} and has since been respecified in 1999, 2014, and 2022.
        Each respecification has brought with it minor changes to the protocol, which has led to significant variation in HTTP/1.1 implementations.

    \subsection{HTTP Server Classes}
\begin{figure}
            \includegraphics[width=\linewidth]{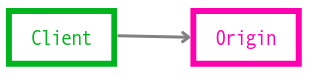}
            \caption{The flow of requests between a client and an origin server}
            \label{fig:origin}
        \end{figure}

\begin{figure}
            \includegraphics[width=\linewidth]{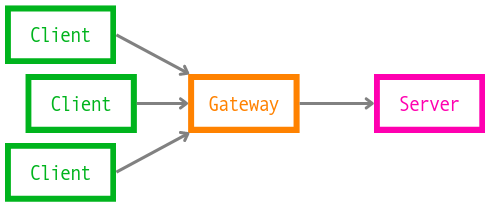}
            \caption{The flow of requests between a client and a gateway server}
            \label{fig:gateway}
        \end{figure}

\begin{figure}
            \includegraphics[width=\linewidth]{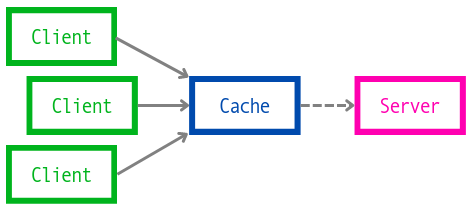}
            \caption{The flow of requests between a client and a cache server. The dotted line represents the fact that not all requests received by the cache are forwarded to its backing server.}
            \label{fig:cache}
        \end{figure}

        RFC 9110 defines many classes of servers in the HTTP/1.1 protocol, of which three are important to this work:
        \begin{enumerate}
            \item \textbf{Origin servers}, which consume incoming requests and produce responses. Figure \ref{fig:origin} shows the flow of requests from a client to an origin server.
            \item \textbf{Gateways}, also known as \textbf{reverse proxies}, which rewrite, normalize, filter, and forward incoming requests and outgoing responses on behalf of a server or group of servers. Figure \ref{fig:gateway} shows the flow of requests from clients through a gateway.
            \item \textbf{Cache servers}, which save responses from other servers, act as origin servers for saved resources and act as intermediary servers otherwise. Figure \ref{fig:cache} shows the flow of requests from clients through a cache.
        \end{enumerate}

        By nature of their roles, cache and gateway servers must act as TLS termination points if TLS is in use.
        Thus, the presence or absence of TLS has no effect on the techniques described in this work.
    
        In order to recognize whether resources are present in the cache, a cache server's interpretation of each incoming request must be shared by the backing server(s).
        Similarly, to effectively modify, filter, and route requests, a gateway server's interpretation of each incoming request must also match that of the backing server(s).
        Thus, gateway servers and cache servers serve very similar roles and are accordingly vulnerable to a similar set of exploitation techniques, such as request smuggling.
        We therefore refer to both gateway servers and cache servers as \textbf{HTTP transducer}s.

    \subsection{Connection Reuse}
\begin{figure}
        \begin{center}
            \includegraphics[width=.75\linewidth]{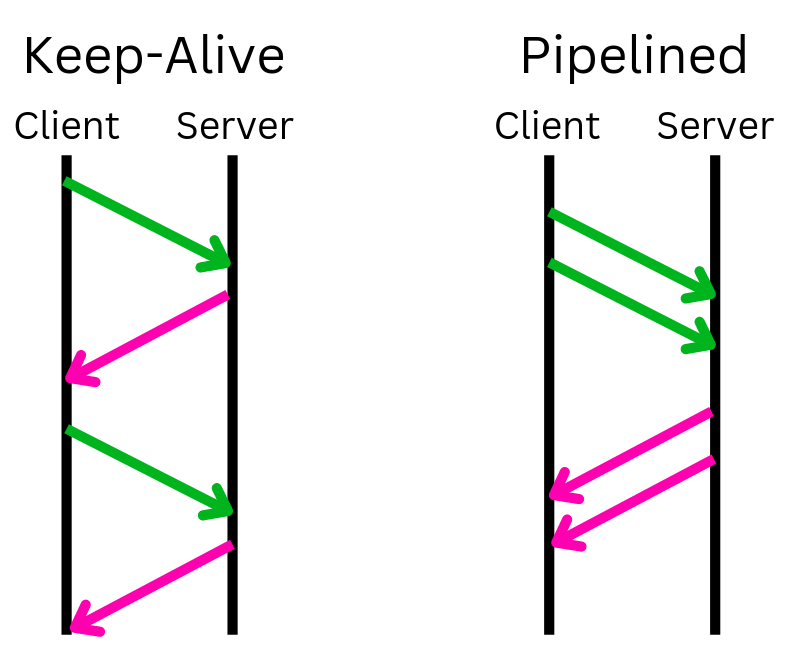}
        \end{center}
        \caption{The two techniques for reusing HTTP/1.1 connections}
        \label{fig:reuse}
        \end{figure}

        To avoid the overhead of repeatedly establishing TCP and TLS connections, the HTTP/1.1 specification offers two features for connection reuse connections: message pipelining and keep-alive connections.
        Pipelined HTTP messages are messages that are sent concatenated together.
        By pipelining requests and responses, a client and server can exchange multiple messages in a single RTT.
        Keep-alive connections are HTTP connections on which requests and responses are repeatedly exchanged over a single persistent connection.
        These techniques are not mutually exclusive and may be used simultaneously.
        Figure \ref{fig:reuse} shows the difference between keep-alive connections and pipelined request sequences.

        Because HTTP transducers often have a fixed set of backing servers, it is efficient to maintain persistent connections to them and forward all incoming requests along those connections.
        In this way, requests from multiple clients may be coalesced on the same connection.
        HTTP/1.1 does not provide an explicit mechanism for associating requests with responses, so senders must associate outgoing requests with incoming responses using order alone.
        It is, therefore, critical to ensure that this association procedure does not fail in a gateway or proxy because it could cause responses, potentially containing sensitive information, to be served to the wrong client.

    \subsection{Message Format}
        Each HTTP/1.1 message consists of three components: the start-line, the headers, and the body~\cite{fielding2022rfc}.

        In a request, the start-line consists of an HTTP method (e.g., \texttt{GET}), a URI (e.g., \texttt{/}), and a protocol version string, (e.g., \texttt{HTTP/1.1}).
        In a response message, the start-line consists of a protocol version string, a status code (e.g., \texttt{404}), and an optional reason string (e.g., \texttt{Page not found}).
        
        A message's headers are colon-separated name-value pairs in which the names are nonempty strings from a subset of the printable ASCII characters, and the values are strings of arbitrary bytes, excluding the null byte, line feed (\texttt{\textbackslash x0a}), and carriage return (\texttt{\textbackslash x0d}).
        Some header names (e.g., \texttt{Content-Type}) have special meaning documented by the HTTP RFCs, but application developers are free to invent their own header names as they see fit.
        Headers are delimited by carriage return-line feed pairs (henceforth referred to as \texttt{CRLF}s) and terminated by the occurrence of two such pairs.

        A message's body is an optional sequence of arbitrary bytes that immediately follows the headers.
        An HTTP/1.1 message body may be encoded either directly or in chunks.
        
        A directly-encoded message body's content consists of the bytes that follow the headers exactly as received.
        In most circumstances, directly encoded message body lengths are specified using the \texttt{Content-Length} header.
        The value of the \texttt{Content-Length} header must be a decimal integer encoded as an ASCII string.
        This value is interpreted as the length of the message body.
        
        A message body that uses the chunked encoding has its content divided into chunks.
        Each chunk consists of a length encoded as a hexadecimal ASCII string, followed by a \texttt{CRLF}, the specified number of data bytes, and another \texttt{CRLF}.
        A chunked message body is terminated by the occurrence of a chunk with length 0.
        To decode a chunked message body, the data bytes are extracted from each chunk and concatenated together.

        In general, parsing a given HTTP message's body is nontrivial due to numerous exceptions to the general principles laid out above.
        For instance, some responses to \texttt{HEAD} and \texttt{CONNECT} requests are defined never to have a message body, even if they contain \texttt{Content-Length} or \texttt{Transfer-Encoding} headers.

    \subsection{HTTP Request Smuggling}
        HTTP request smuggling occurs when an HTTP transducer's interpretation of an incoming request stream differs substantially from its backing server's interpretation of the corresponding forwarded request stream such that at least one request in the stream is not invisible to the transducer but is interpreted by the backing server.
        When this occurs, we say the invisible request has been ``smuggled."
        Most commonly, request smuggling is caused by discrepancies in message body parsing between a transducer and its backing server, but it can also be caused by header and start-line parsing discrepancies.

        When a cache server falls victim to HTTP request smuggling, it can no longer correctly associate requests to responses and may, therefore, cache the response to the smuggled request under the wrong key.
        This is known as web cache poisoning and can be used both for denial of service and to leak sensitive data from web services~\cite{cpdos}.

        When a gateway server falls victim to HTTP request smuggling, its normalizations and filtering rules will not be applied to the smuggled request.
        Access controls implemented by a gateway server may be bypassed in this manner.

        When transducers are reusing connections to send requests from multiple different clients, request smuggling may cause all or part of a response intended for one client to be served to another client.
        This is known as HTTP desync~\cite{desync} and can enable attackers to steal sensitive data from responses intended for other clients.

    \subsection{CPDoS}
        It is also possible to poison caches without needing to smuggle entire requests.
        For example, if a cache server sees a request for \texttt{/}, and its backing server sees a request for \texttt{/{\textbackslash}t}, the cache server will store the server's response (likely a \texttt{404}) under the cache key \texttt{/}.
        Other visitors to the site are then met with a \texttt{404} until the cache timeout expires.
        This is known as cache-poisoned denial-of-service (CPDoS).

\section{Related Work}

This paper builds on prior work in fuzzing, differential testing, and HTTP request smuggling.
We discuss the closest work in more detail.

    \subsection{T-Reqs and FRAMESHIFTER}
        T-Reqs is a differential fuzzer that searches for request smuggling vulnerabilities in HTTP transducers~\cite{treqs}.
        FRAMESHIFTER is a differential fuzzer that searches for HTTP/2 downgrading discrepancies in HTTP transducers.
        Each of these uses an HTTP grammar to generate valid parse trees, which are then mutated, serialized, and forwarded to a collection of HTTP transducers.
        The transducers then forward the requests to a so-called ``feedback server," which is described as depositing the received requests in a database.\footnote{We also note that this claim is not supported by the T-Reqs codebase. In reality, T-Reqs's ``feedback server" is an echo server similar to HDiff's, but it echoes only what it interprets as the received message body.}
        The forwarded requests are then re-parsed by a ground-truth HTTP parser, and discrepancies between their parsed message bodies are reported to the user.

        This approach hinges on a few assumptions that do not always hold in practice.
        First, the ground truth parser's interpretation of the transducer's output may differ from the transducer's own interpretation.
        When this occurs, it is impossible for the ground truth parser to recognize exploitable transducer behavior accurately.
        Second, by nature of being transducer-oriented, T-Reqs is able to detect only discrepancies for which the fault falls upon transducer(s).
        Later work demonstrated that the most severe parsing discrepancies exist within origin servers~\cite{grenfeldt}.
        For example, we have discovered HTTP request smuggling vulnerabilities that do not rely on the existence of bugs in transducers, and we describe them in more detail in Section \ref{sec:findings}.
        Third, T-Reqs's input generator is purely random; there is no feedback mechanism whatsoever.
        This causes many repeated results and makes combing through fuzzer output time-consuming and difficult.
        Fourth, T-Reqs's input generator and mutator are targeted toward generating single requests and, therefore, cannot discover request smuggling bugs that require multiple message exchanges using keep-alive connections.

    \subsection{HDiff}
        HDiff is a differential fuzzer for HTTP transducers and origin servers that searches for bugs involving inconsistent interpretation of HTTP requests.
        HDiff generates requests according to a grammar extracted from RFC 7230, mutates these requests very much as T-Reqs does, and tags the mutants with a unique ID that is contained within the request's URI query.
        Instead of collecting transducer output directly with a feedback server, it is reflected through an ``echo server," which responds to any received bytes with a response containing those same bytes encapsulated within the response body.
        To collect output from origin servers, each is equipped with a small script in either PHP or Java, which echoes back the message body, URI, and host header from each received request.
        Additionally, some of the same information is collected from the transducers by making use of their logging capabilities.

        This approach improves on T-Reqs by examining more than transducer output alone but still assumes that request smuggling can be detected by examining only a fraction of a given request's parse tree.
        T-Reqs itself demonstrated that even discrepancies in HTTP versions and methods can be used for request smuggling~\cite{treqs}.
        HDiff, like T-Reqs, is also limited in its fuzzing capability.
        Since HDiff has no feedback mechanism, all generated inputs that do not cause discrepancies are discarded.
        This includes inputs that are only a single bitflip away from causing a discrepancy.

    \subsection{Differential Testing}

    Differential testing, or differential fuzzing, is the process of systematically comparing programs that are intended to implement the same protocol~\cite{differential-testing}.
    Researchers have applied differential testing to various use cases where multiple implementations must follow similar specifications.
    For example, in SFADiff, Argyros et al.~\cite{10.1145/2976749.2978383} explore how grammar learning techniques can be used in identifying program state machines and how these learned state machines can be compared across implementations.
    
    DifuzzRTL~\cite{9519470} performs differential testing across CPU ISA and RTL simulations to establish if there are bugs in either. QDiff~\cite{9678792} tested different quantum software stacks to find implementation errors.
    There have also been repeated attempts to perform differential testing across browsers and their JavaScript engines~\cite{9402086, 10.1145/3510003.3510044, 10.1145/3548606.3560624}.
    Sorniotti et al.~\cite{sorniotti2023go} explored the question of what discrepancies may exist between C/C++ and native implementations of Go libraries.
    Li et al.~\cite{10.1145/3582016.3582053} used different compilers on the same source code and then compared the binaries to test the hypothesis that each compiler would handle undefined behavior differently.
    Finally, Yang et al.~\cite{yang2021finding} demonstrated how differential fuzzing can be used to find consensus bugs in Ethereum implementations.
    There have been numerous comparisons of TLS implementations as well~\cite{10.1145/3355048, 10.1145/3510416, 8070382}.
    
    Our differential fuzzer design builds on several design choices made in NEZHA~\cite{nezha}.
    We use a differential coverage metric first described in NEZHA: fine path $\delta$-diversity.
    We also use it to implement a genetic fuzzer, similar to how it is employed in NEZHA.

\section{HTTP Garden}

\label{sec:http-garden}

\begin{table*}[t!]
    \centering
        \resizebox{\textwidth}{!}{
    \begin{tabular}{l|c|c|l|c}
        \textbf{Server} & \textbf{Traced}& \textbf{Internal/External} & \textbf{Source} & \textbf{Vulnerabilities Reported} \\\hline
Aiohttp	& Yes& Internal & \url{https://docs.aiohttp.org/en/stable/} & 10 \\
Apache	& Yes& Internal & \url{https://httpd.apache.org/} & 2 \\
Bun	& No& Internal & \url{https://bun.sh/} & 2\\
CherryPy& No& Internal & \url{https://cherrypy.dev/} & 1 \\
Daphne & Yes& Internal & \url{https://github.com/django/daphne} & 2 \\
Deno & No& Internal & \url{https://deno.com/} & 0 \\
fasthttp & No& Internal & \url{https://github.com/valyala/fasthttp} & 3 \\
Go net/http&No& Internal & \url{https://pkg.go.dev/net/http} & 4 \\
Gunicorn&No& Internal & \url{https://gunicorn.org/} & 6 \\
H2O&Yes& Internal & \url{https://h2o.examp1e.net/} & 3 \\
Hyper&No& Internal & \url{https://hyper.rs/} & 1 \\
Hypercorn&No& Internal & \url{https://pypi.org/project/Hypercorn/} & 0 \\
Jetty&No& Internal & \url{https://eclipse.dev/jetty/} & 0 \\
libevent&No& Internal & \url{https://libevent.org/} & 3 \\
libsoup&No& Internal & \url{https://libsoup.org/libsoup-3.0/index.html} & 0 \\
Lighttpd&Yes& Internal & \url{https://www.lighttpd.net/} & 1 \\
Microsoft IIS & No & External & \url{https://www.iis.net/} & 0 \\
Mongoose&Yes& Internal & \url{https://mongoose.ws/} & 9 \\
Nginx&Yes& Internal & \url{https://www.nginx.com/} & 1 \\
Node.js&No& Internal & \url{https://nodejs.org/} & 3 \\
OpenLiteSpeed&No& Internal & \url{https://github.com/litespeedtech/openlitespeed} & 10 \\
OpenBSD httpd & No & External & \url{https://man.openbsd.org/httpd.8} & 2 \\
Passenger&No& Internal & \url{https://www.phusionpassenger.com/} & 2 \\
Puma&No& Internal & \url{https://puma.io/} & 3 \\
Tomcat&No& Internal & \url{https://tomcat.apache.org/} & 0 \\
Tornado&Yes& Internal & \url{https://www.tornadoweb.org/} & 7 \\
uhttpd&Yes& Internal & \url{https://git.openwrt.org/project/uhttpd} & 0 \\
Unicorn&No& Internal & \url{https://yhbt.net/unicorn/README.html} & 0 \\
Uvicorn&Yes& Internal & \url{https://www.uvicorn.org/} & 0 \\
Waitress&Yes& Internal & \url{https://docs.pylonsproject.org/projects/waitress/en/stable/} & 2 \\
WEBrick&No& Internal & \url{https://github.com/ruby/webrick} & 3 \\
Werkzeug&No& Internal & \url{https://werkzeug.palletsprojects.com/en/3.0.x/} & 1 \\\hline
        \textbf{Proxy} & \textbf{Traced}& \textbf{Internal/External} & \textbf{Source} & \textbf{Vulnerabilities Reported} \\\hline
Apache Proxy & No & Internal & \url{https://httpd.apache.org/} & 0 \\
Apache Traffic Server & No & Internal & \url{https://github.com/apache/trafficserver} & 6 \\
Caddy Proxy & No& Internal  & \url{https://caddyserver.com/} & 0 \\
H2O Proxy & No & Internal & \url{https://h2o.examp1e.net/} & 2 \\
HAProxy & No & Internal & \url{https://www.haproxy.org/} & 3 \\
nghttpx & No & Internal & \url{https://github.com/nghttp2/nghttp2} & 1 \\
Nginx Proxy & No& Internal & \url{https://www.nginx.com/} & 0 \\
OpenLiteSpeed Proxy & No& Internal & \url{https://github.com/litespeedtech/openlitespeed} & 5 \\
OpenBSD relayd & No & External  & \url{https://man.openbsd.org/relayd.8} & 8 \\
Pound & No& Internal & \url{https://maucher-online.com/pound/} & 4\\
Squid & No& Internal & \url{https://www.squid-cache.org/} & 1 \\
Varnish HTTP Cache & No& Internal & \url{https://varnish-cache.org/} & 1 \\ \hline
    \end{tabular}}
    \caption{The set of HTTP implementations in the HTTP Garden testbench setup. All the open-source implementations use the latest commit available; for the Microsoft IIS server, we use the latest release.}
    \label{tab:garden}
\end{table*}

        The HTTP Garden is, at its core, a collection of HTTP origin servers and transducers that are configured for compatibility. The full collection of servers supported is available in Table~\ref{tab:garden}. Most servers in the Garden run within local Docker containers, but some servers, such as Akamai GHost and Google Cloud Classic Application Load Balancer, cannot be run locally, and must therefore be hosted externally to the Garden. A full list of external, cloud-based transducers we included in our setup is available in Table~\ref{tab:proxyendpoints}.

        There are three classes of servers in the HTTP Garden: origin servers, transducers, and echo servers.

        Origin servers respond to each incoming request with a response body containing a JSON object with the following fields: \texttt{body}, \texttt{uri}, \texttt{headers}, and \texttt{version}.
        These correspond to the parsed contents of the incoming request.
        Each of these fields is encoded in Base64 to obviate the need for escaping strings.

\begin{figure}
            \begin{center}
            \includegraphics[width=\linewidth]{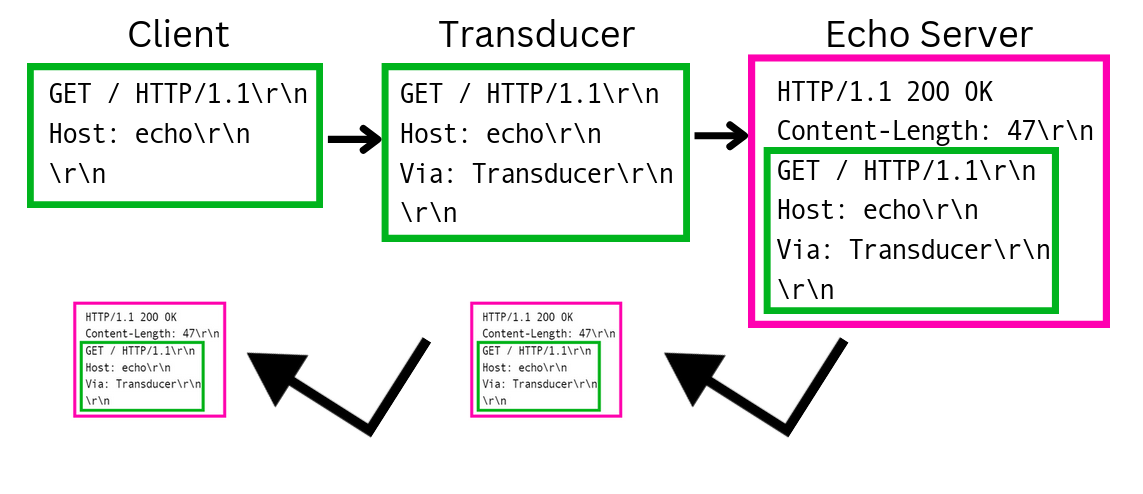}
            \end{center}
            \caption{A request being routed through a transducer and echo server.}
            \label{fig:echo}
        \end{figure}

        Transducers route all incoming requests to an echo server.
        The transducers then route the echo server's response back to the client.
        Figure~\ref{fig:echo} shows the route that a request takes through a transducer in the Garden.

\begin{table*}[h]
    \centering
        \resizebox{\textwidth}{!}{
    \begin{tabular}{l | c | c | c | c | c}
        \textbf{Service} & \textbf{Requires SSL to Endpoints} & \textbf{Works without SSL?} & \textbf{Configured as} & \textbf{Considered in T-Reqs} & \textbf{Vulnerabilities} \\ \hline
        Cloudflare & Yes & Yes & CDN Proxy & Yes & 0\\
        AWS CloudFront & Yes & No & Load Balancer & Yes & 0 \\
        Akamai & No & No & CDN Proxy & Yes & 4 \\
        Fastly & No & Yes & CDN Proxy & No & 0 \\
        Azure & No & Yes & CDN Proxy & No & 1 \\
        Google Classic & No & Yes & Load Balancer & No & 2 \\
        Google Global & No & Yes & Load Balancer & No & 0 \\
        \hline
    \end{tabular}}
    \caption{We set up these services as proxies or load balancers in the HTTP Garden}
    \label{tab:proxyendpoints}
\end{table*}

        \paragraph{Echo Servers.} Echo servers are simple TCP servers that read incoming data until a timeout occurs, then respond with an HTTP \texttt{200 OK} response containing that data in its message body.
        This process is then repeated until the timeout occurs with no data received, upon which the TCP connection is closed.
        Because the echo servers respond to data as it arrives, a client can reconstruct the relative timing of byte stream elements as they were sent from the transducer. %
        This allows the echo server's client to detect, for example, if a transducer is un-pipelining request sequences.

        The abundance of supported servers in the HTTP Garden demonstrates the ease with which new servers can be integrated.

        We employ two echo servers: one that runs locally to service requests from transducers that run in local Docker containers and another that runs on a publicly accessible virtual-private server (VPS) to service requests from external transducers.

\section{Coverage-Guided Differential Fuzzing}

\label{sec:diff-fuzz}

    \subsection{Fuzzing}
    
        The Garden revolves around a simple fuzzing loop.
        Beginning with a small corpus of HTTP requests, inputs are sent to each server, and their responses and coverage are collected.
        If an input causes origin servers' responses to differ, that input is evaluated for \textbf{meaningfulness} and \textbf{durability}.
        It is reported as a result if it is both meaningful and durable.
        If an input that does not cause a discrepancy is deemed worthy of mutation, it is added to the mutation queue to create the next generation of inputs.
        This loop continues for a user-configurable number of generations and with a user-configurable generation size.

        \begin{figure}[t]
            \includegraphics[width=\linewidth]{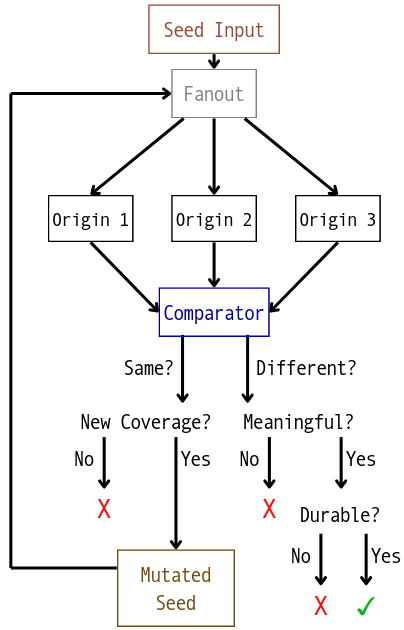}
            \caption{The HTTP Garden Fuzzer}
        \end{figure}

    \subsection{Feedback}

        Some targets are instrumented for the collection of control flow coverage.
        For C and C++ targets, this is accomplished using AFL++'s \texttt{afl-cc} compiler wrapper.
        This is accomplished with the \texttt{python-afl} PyPI package for Python targets.
        For the remainder of the targets, control flow coverage is not collected, but observed differentials involving these targets are still reported to the user.
    
        Traditionally, fuzzers have relied on the assumption that targets exit after processing input.
        However, HTTP servers do not exit after servicing each request.
        We made a slight (38 LoC) modification to the AFL++ source to accommodate this fact.
        In particular, we introduce two signal handlers to \texttt{afl-showmap}.
        When \texttt{afl-showmap} receives \texttt{SIGUSR1}, it writes the current state of the coverage map into a temporary file, \texttt{/tmp/trace}.
        When \texttt{afl-showmap} receives \texttt{SIGUSR2}, it clears the coverage map.

        Control flow coverage from the processing of a given request can thus be collected from the servers by sending the ``clear map" signal, sending the request, receiving the response, sending the ``write map" signal, and reading the table from the temporary file.

        Because the coverage collection effectively begins before the request is sent and ends after the response is received, there is sometimes a small amount of extra coverage collected.
        This introduces some false positives into the mutation queue but will never introduce a false negative due to the parent selection process.
        We deem this to be an acceptable trade-off between accuracy, performance, and convenience compared to desocketing approaches.

    \subsection{Parent Selection}

        Inputs are selected for mutation according to the fine path $\delta$-diversity metric~\cite{nezha}.
        When an input causes a never-before-encountered $\delta$-diversity tuple, it is always selected for mutation.
        When an input causes a discrepancy, that input is never selected for mutation, even if its $\delta$-diversity tuple is new.
        This prevents the fuzzer from churning out endless variations on the same underlying bugs.

    \subsection{Mutations}

        Although mutations have operated upon individual byte strings in previous work~\cite{treqs,hdiff}, the mutator in the HTTP Garden operates upon sequences of byte strings, which we refer to as ``byte streams."
        To send a byte stream to a server, the Garden sends an element, then receives bytes until a timeout occurs, and repeats this process until all elements have been sent and all server output collected.
        This enables the fuzzer to better discover discrepancies related to connection reuse.
        
        The Garden supports three types of mutations: byte mutations, stream mutations, and grammar-based mutations.
        Byte mutations add, replace, or remove bytes from one of the elements of the byte stream.
        Stream mutations add, replace, combine, or remove elements from the byte stream.
        Grammar-based mutations modify the higher-level constructs represented by the strings in the stream.
        A grammar-based mutation might, for example, replace an HTTP method contained within a stream element with another valid HTTP method.
        Of course, this requires that the stream element be an HTTP request with a recognizable method and, therefore, cannot be applied to arbitrary stream elements.
        We circumvent this problem using an extremely lenient HTTP parser for applying mutations.
        This allows us to repeatedly mutate requests using grammar-based mutations, improving upon prior work, which was limited to applying grammar-based mutations only to freshly generated inputs.

    \subsection{Reporting Criteria}

        Due to the myriad optional portions of the HTTP RFCs, the HTTP Garden's fuzzer needs to selectively ignore many uninteresting discrepancies during a typical run.
        This is handled by reporting only meaningful, durable discrepancies.

        \paragraph{Meaningfulness.} Meaningfulness is determined by a comparison operation that considers the known behavior of the Garden's targets with respect to optional portions of the RFCs.
        This allows the fuzzer to avoid reporting inputs that cause discrepancies due to, for example, servers accepting HTTP/0.9 in addition to HTTP/1.1.
        To filter out these acceptable discrepancies, a record of each server's permissible parsing quirks must be generated before starting the fuzzer.
        The Garden includes a script that dynamically generates such a record for each configured server by sending a sequence of requests capable of diagnosing common permissible HTTP/1.1 implementation discrepancies.
        After removing the inputs that cause acceptable discrepancies, the remainder are declared meaningful.
        \paragraph{Durability.} A discrepancy-inducing input $I$ is durable if there exists a transducer $T$ in the Garden such that the result of sending $I$ through $T$ also induces a discrepancy between at least one pair of origin servers in the Garden.
        This requirement exists because discrepancy-inducing inputs that are normalized into benign inputs by transducers are not useful for attacking server chains.

        If a discrepancy-inducing input is meaningful and durable, it is reported to the user as a fuzzing result.

    \subsection{REPL}

        Once the fuzzer is done executing, each durable discrepancy-inducing input $I$ is mapped to a symmetric $n \times n$ boolean discrepancy matrix $M(I)$, where $n$ is the number of origin servers in the Garden, and $M(I)_{i,j} = 1$ if and only if servers $i$ and $j$ disagree about the interpretation of the input $I$.
        Durable discrepancy-inducing inputs that map to the same discrepancy matrix are reported to the user together because they likely trigger the same underlying bug(s).

        After the results have been reported, users can interact with the target servers and the fuzzer's output using the Garden's REPL.
        The supported interactions include applying further mutations to the fuzzer's output, crafting new inputs to be sent to the origin servers and transducers, directly observing and comparing origin server parse trees and the effects of transducers, and computing and displaying discrepancy matrices for user-provided inputs.
        This allows users to pick up where the fuzzer left off, simplifying and improving the fuzzer's output until mere discrepancy-inducing inputs are polished into practical exploits.

\section{Findings}

\label{sec:findings}

    We found 122 unique parsing discrepancies using the HTTP Garden, of which 68 have been patched. A detailed list of these discrepancies can be found in our GitHub repository.

    \subsection{Integer Parsing Attacks}
        HTTP relies on two different types of length fields: \texttt{Content-Length} values and chunk sizes.
        A \texttt{Content-Length} value must consist of a nonempty string of ASCII decimal digits.
        A chunk size must consist of a nonempty string of ASCII hexadecimal digits.
        Surprisingly, an integer parser that matches these criteria does not exist in the C or Python standard libraries.
        The C standard library integer parsers (e.g., \texttt{strtol}) allow digit-separating underscores, leading and trailing whitespace, \texttt{+} and \texttt{-} prefixes, and \texttt{0x} prefixes on hexadecimal numbers.
        The Python standard library integer parser, \texttt{int}, adds to that list digit-separating underscores.
        This addition was introduced in Python 3.6, so servers that updated without reading the release notes may have inadvertently introduced new bugs to their systems~\cite{pep515}.

        The \texttt{strtol}-family integer parsers also allow for radix inference by specifying a radix of \texttt{0}.
        Most notably, if the input begins with \texttt{0}, it is interpreted in octal.
        One popular HTTP server, LiteSpeed, uses the radix inference mode of \texttt{strtoll} to parse \texttt{Content-Length} values.
        Thus, when a transducer forwards a request with a \texttt{0}-prefixed \texttt{Content-Length} value, LiteSpeed will misinterpret the request body.
        While some transducers normalize out leading \texttt{0}s in \texttt{Content-Length} values, many do not, and the RFCs do not require either behavior.
        Among those that do not normalize leading \texttt{0}s are Apache Traffic Server, nghttpx, Pound, Squid, Varnish, Google Cloud Classic Application Load Balancer, and OpenBSD relayd.
        HAProxy began normalizing leading \texttt{0}s in \texttt{Content-Length} values when we mentioned to the maintainer that normalizing leading \texttt{0}s would close a vulnerability in LiteSpeed, even though HAProxy itself complies with the RFCs.
        This discrepancy can be used to smuggle requests to instances of LiteSpeed deployed behind any of these transducers.

        \begin{figure}[t]
        \begin{subfigure}[a]{\linewidth}
        \begin{lstlisting}[frame=single]
<@\textcolor{red}{GET / HTTP/1.1}@>\r\n
<@\textcolor{blue}{Content-Length: 0200}@>\r\n
\r\n
<@\textcolor{magenta}{AAAAAAAAAAAAAAAAAAAAAAAAA}@>
<@\textcolor{magenta}{AAAAAAAAAAAAAAAAAAAAAAAAA}@>
<@\textcolor{magenta}{AAAAAAAAAAAAAAAAAAAAAAAAA}@>
<@\textcolor{magenta}{AAAAAAAAAAAAAAAAAAAAAAAAA}@>
<@\textcolor{magenta}{AAAAAAAAAAAAAAAAAAAAAAAAAAAA}@>
<@\textcolor{magenta}{GET /.ssh/id\_rsa HTTP/1.1{\textbackslash}r{\textbackslash}n}@>
<@\textcolor{magenta}{Content-Length: 56{\textbackslash}r{\textbackslash}n}@>
<@\textcolor{magenta}{{\textbackslash}r{\textbackslash}n}@>
<@\textcolor{magenta}{AAAAAAAAAAAAAAAAAAAAAAA}@>
<@\textcolor{red}{GET / HTTP/1.1}@>\r\n
<@\textcolor{blue}{Host: whateva}@>\r\n
\r\n
        \end{lstlisting}
        \caption{Transducer's interpretation of the attack payload.}
        \end{subfigure}

        \begin{subfigure}[b]{\linewidth}
        \begin{lstlisting}[frame=single]
<@\textcolor{red}{GET / HTTP/1.1}@>\r\n
<@\textcolor{blue}{Content-Length: 0200}@>\r\n
\r\n
<@\textcolor{magenta}{AAAAAAAAAAAAAAAAAAAAAAAAA}@>
<@\textcolor{magenta}{AAAAAAAAAAAAAAAAAAAAAAAAA}@>
<@\textcolor{magenta}{AAAAAAAAAAAAAAAAAAAAAAAAA}@>
<@\textcolor{magenta}{AAAAAAAAAAAAAAAAAAAAAAAAA}@>
<@\textcolor{magenta}{AAAAAAAAAAAAAAAAAAAAAAAAAAAA}@>
<@\textcolor{red}{GET /.ssh/id\_rsa HTTP/1.1}@>\r\n
<@\textcolor{blue}{Content-Length: 56}@>\r\n
\r\n
<@\textcolor{magenta}{AAAAAAAAAAAAAAAAAAAAAAA}@>
<@\textcolor{magenta}{GET / HTTP/1.1{\textbackslash}r{\textbackslash}n}@>
<@\textcolor{magenta}{Host: whateva{\textbackslash}r{\textbackslash}n}@>
<@\textcolor{magenta}{{\textbackslash}r{\textbackslash}n}@>
        \end{lstlisting}
        \caption{LiteSpeed's interpretation of the attack payload.}
        \end{subfigure}
        \caption{Request smuggling via bad integer parsing.}
        \end{figure}
        
        We also found numerous integer parsing attacks in the Python HTTP ecosystem.
        Origin servers using the \texttt{int} integer parser were vulnerable to request smuggling when paired with transducers that accept and forward invalid chunk sizes.
        These transducers typically interpret such invalid chunk sizes as equivalent to their longest valid prefix.
        One such transducer is Apache Traffic Server (ATS).
        Therefore, when ATS receives a request containing a chunk size of \texttt{0\_ff}, for example, it interprets that chunk size as being equivalent to \texttt{0}, signaling the termination of the chunked message body and the beginning of the next request.
        However, because it forwards the invalid chunk size as-is, an HTTP server that ignores digit-separating underscores will see the same chunk as having a length of 255.
        Many Python HTTP servers, including AIOHTTP, Gunicorn, and Tornado, ignored digit-separating underscores due to the confusing behavior of the Python \texttt{int} parser.
        The ATS maintainers have alerted us that they are currently developing a patch for their half of this vulnerability.
        All three of the aforementioned Python HTTP servers have now patched this behavior since we reported it to them.

        This same ATS bug is also exploitable to smuggle requests to servers that use \texttt{strtol}-family integer parsers, even when the radix is given explicitly.
        This is because these integer parsers ignore ``\texttt{0x}" prefixes on their input if there is an explicitly provided radix of 16.
        Origin servers in this category include CherryPy, Libevent, Libsoup, and OpenWrt uhttpd.
        We are currently in the process of reporting these vulnerabilities.

    \subsection{Chunk Parsing Attacks}
        Each HTTP message body chunk comprises two parts: a size line and a data line.
        Length lines may optionally contain a ``chunk extension," which is a key-value pair separated by ``\texttt{=}" that immediately follows a chunk's size and is ignored by nearly all modern HTTP implementations.
        Chunk extensions are separated from chunk sizes with a ``\texttt{;}" character, with optional whitespace permitted on either side.
    
        The HTTP RFCs state that both types of chunk lines must be delimited by \texttt{CRLF}.
        However, using the HTTP Garden, we discovered that Node.js allows chunk lines to be terminated by bare \texttt{CR}.
        Further, we discovered that this discrepancy is durable because Apache Traffic Server, Google Cloud Classic Application Load Balancer, and Akamai GHost all accept and forward \texttt{CR} bytes within the optional whitespace before the ``\texttt{;}" that delimits a chunk extension.
        By iterating on this discrepancy in the Garden, we were able to develop payloads to execute request smuggling against any of these transducers when Node.js is used as their backing server.

        \begin{figure}[t]
        \begin{subfigure}[a]{\linewidth}
        \begin{lstlisting}[frame=single]
<@\textcolor{red}{POST / HTTP/1.1}@>\r\n
<@\textcolor{blue}{Transfer-Encoding: chunked}@>\r\n
\r\n
<@\textcolor{orange}{2}@>\r\r<@\textcolor{magenta}{;a}@>\r\n
<@\textcolor{orange}{02}@>\r\n
<@\textcolor{magenta}{2d}@>\r\n
<@\textcolor{orange}{0}@>\r\n
\r\n
<@\textcolor{red}{DELETE / HTTP/1.1}@>\r\n
<@\textcolor{blue}{Content-Length: 23}@>\r\n
\r\n
<@\textcolor{magenta}{0{\textbackslash}r{\textbackslash}n}@>
<@\textcolor{magenta}{{\textbackslash}r{\textbackslash}n}@>
<@\textcolor{magenta}{GET / HTTP/1.1{\textbackslash}r{\textbackslash}n}@>
<@\textcolor{magenta}{{\textbackslash}r{\textbackslash}n}@>
        \end{lstlisting}
        \caption{Node.js's interpretation of the attack payload}
        \end{subfigure}
        \begin{subfigure}[b]{\linewidth}
        \begin{lstlisting}[frame=single]
<@\textcolor{red}{POST / HTTP/1.1}@>\r\n
<@\textcolor{blue}{Transfer-Encoding: chunked}@>\r\n
\r\n
<@\textcolor{orange}{2}@>\r\r;a\r\n
<@\textcolor{magenta}{02}@>\r\n
<@\textcolor{orange}{2d}@>\r\n
<@\textcolor{magenta}{0{\textbackslash}r{\textbackslash}n}@>
<@\textcolor{magenta}{{\textbackslash}r{\textbackslash}n}@>
<@\textcolor{magenta}{DELETE / HTTP/1.1{\textbackslash}r{\textbackslash}n}@>
<@\textcolor{magenta}{Content-Length: 23{\textbackslash}r{\textbackslash}n}@>
<@\textcolor{magenta}{{\textbackslash}r{\textbackslash}n}@>
<@\textcolor{magenta}{0{\textbackslash}r{\textbackslash}n}@>
<@\textcolor{magenta}{{\textbackslash}r{\textbackslash}n}@>
<@\textcolor{red}{GET / HTTP/1.1}@>\r\n
\r\n
        \end{lstlisting}
        \caption{ATS, Akamai, and Google Cloud's interpretation of the attack payload}            
        \end{subfigure}
        \caption{Request smuggling via bad chunk delimiter parsing.}
        \end{figure}
        
        We could also bypass arbitrary access controls implemented by the three transducers (e.g., URI filtering).

        In response to our discovery, Node.js patched their HTTP parser to fix this bug in release 21.2.0.
        Apache Traffic Server has not yet released a fix, but a maintainer has stated that one is being developed.
        Google and Akamai each rolled out mitigations for this attack.
        Akamai began rejecting all requests containing \texttt{CR} bytes within the relevant whitespace field.
        Google began stripping out chunk extensions and their preceding optional whitespace from all incoming requests.
        Unfortunately, neither of these mitigations prevents the attack.
        Through further exploration in the Garden, we discovered that Google's mitigation does not strip \texttt{CR} bytes that immediately follow chunk sizes but are not followed by a ``\texttt{;}".
        We also discovered that Akamai's mitigation does not strip \texttt{CR} bytes from the optional whitespace that follows the ``\texttt{;}".
        We have reported the shortcomings of both of these mitigations, and Akamai has assured us that they are working on a better patch.
        We are waiting to hear back from Google.

        We discovered another exploitable chunk line parsing discrepancy in the Puma HTTP server.
        Puma allowed chunked message bodies to be terminated with \texttt{CRLF} followed by any two bytes.
        When those two bytes are a single ASCII alphanumeric character followed by a ``\texttt{:}", and then the string ``\texttt{GET /evil HTTP/1.1}", a standards-compliant transducer will correctly interpret this as a trailer field with a single-letter name and a value of ``\texttt{GET /evil HTTP/1.1}".
        However, Puma will interpret the letter and the ``\texttt{:}" as the end of the chunked message body and interpret ``\texttt{GET /evil HTTP/1.1}" as the beginning of a new request.
        This bug is, therefore, usable for smuggling requests to Puma past any transducer that both forwards trailer fields and does not add a whitespace after the ``\texttt{:}" in trailer fields.
        Apache Traffic Server, Pound, and OpenBSD relayd are examples of such transducers.
        This vulnerability was fixed in Puma, and a CVE was assigned shortly after we reported it.

    \subsection{Field Line Parsing Attacks}
        The HTTP/1.1 RFCs specify that header field lines must be delimited by either \texttt{LF} or \texttt{CRLF}.
        However, some HTTP implementations, including the Python standard library, accept bare \texttt{CR} as a header line delimiter.
        The RFCs also require that transducers not forward bare \texttt{CR} within header values, but not all transducers follow this rule.
        Notably, Google Cloud Classic Application Load Balancer does not follow this rule and will accept and forward header values containing bare \texttt{CR} without treating it as a line ending.
        When servers using the Python standard library's HTTP implementation are deployed behind Google's load balancer, this discrepancy allows attackers to insert headers directly into the Python server's interpretation of the request without those headers having been interpreted by Google.
        For example, an attacker who wanted to sneak a \texttt{Content-Length: 10} header past Google to a Python-powered web server could do so by sending a \texttt{Whatever: whatever{\textbackslash}rContent-Length: 10{\textbackslash}r{\textbackslash}n} header.

        We also discovered a header parsing attack in OpenBSD's gateway server, relayd.
        When relayd received a request containing either a null byte or bare \texttt{LF} within a header value, it would concatenate the contents of that header's value into the previous header value.
        Of particular note is that this concatenation occurs after relayd has validated and interpreted the value of the \texttt{Content-Length} header.
        Thus, attackers can arbitrarily modify the value of the \texttt{Content-Length} header that relayd forwards, without risking relayd rejecting the message for being malformed.
        While the vast majority of request smuggling bugs rely on bugs in multiple participants of the message exchange, this bug on its own can be used to smuggle requests past relayd to any backend HTTP server.
        Notably, this allows for request smuggling from one instance of relayd to another instance of relayd.
        This is the first such vulnerability of which we are aware.
        This bug was fixed by the OpenBSD team soon after our report.

    \subsection{Field Value Parsing Attacks}
        The HTTP/1.1 RFCs specify that receivers of list-valued header fields must ignore a reasonable number of empty list elements.
        However, the standards also forbid senders from emitting messages with empty list elements.
        The \texttt{Transfer-Encoding} header is list-valued and therefore is subject to these requirements.
        However, since these requirements are not encoded in the HTTP/1.1 ABNF grammar, it is to be expected that implementations differ in their adherence to these rules.
        For instance, Gunicorn, Mongoose, and Passenger all treated the \texttt{,chunked} transfer coding as distinct from the \texttt{chunked} transfer coding, in violation of the RFCs until our reports.
        Correspondingly, until our reports, Azure CDN and nghttpx both treated \texttt{,chunked} as equivalent to \texttt{chunked}, but did not normalize away the leading ``\texttt{,}".
        Thus, when any of the aforementioned origin servers are deployed behind either of the aforementioned transducers, requests can be smuggled.
        
    \subsection{Denial of Service Attacks}
        We also discovered two severe denial-of-service attacks using the HTTP Garden.
        
        The first was in the web server Cesanta Mongoose, which is targeted at embedded systems.
        We found that sending requests with negative \texttt{Content-Length} values could force the server into an infinite busy loop, endlessly parsing the same request.
        This was possible because Mongoose's first-pass parser skips message bodies without validating that their lengths are nonnegative.
        Thus, by sending the right negative \texttt{Content-Length} value, you can skip the parser's ``read head" back to the beginning of the input buffer.
        This bug was patched, and a CVE was assigned to it after our report.

        The second DoS we encountered was in OpenBSD httpd.
        This web server replaced NGINX in the default install of OpenBSD in 2015.
        Although OpenBSD is known for its secure programming practices, it took only a short fuzzing run to crash the server using the HTTP Garden.
        This occurred because of the dereference of an unchecked pointer argument in the server's implementation of FastCGI, which was null when malformed messages were pipelined after valid messages with chunked bodies.
        This vulnerability could be used to crash any instance of OpenBSD httpd that used FastCGI.
        The bug was fixed after our report, and we intend to have a CVE number assigned to it.

\section{Discussion}

    \paragraph{Comparing the HTTP Garden to HDiff and T-Reqs} Comparing differential fuzzers is difficult because determining whether two fuzzers can discover the same bugs requires either human evaluation or the ability to automatically map discrepancies to their underlying bugs.
    Neither of these is feasible in the context of HTTP/1.1 due to the volume of parser discrepancies.
    We therefore argue for the value of the HTTP Garden not by comparing its output directly to its predecessors but by demonstrating that users of the HTTP Garden are capable of discovering new bugs that were missed by T-Reqs and HDiff.

    \paragraph{What's Missing from the Standards?} The HTTP/1.1 RFCs define an ABNF grammar for HTTP/1.1 messages.
    Unfortunately, that grammar specifies only the format of messages a sender can produce.
    The rules that specify which messages receivers are permitted to accept are written in plain English across multiple RFCs.
    Further, 
    If the HTTP working group wants to maintain different grammars for what should be accepted and what should be produced, then this should be made explicit, and two formal grammars should be provided.
    Ideally, those formal grammars would use a format that can encode length fields, which is beyond the capabilities of ABNF.

    \paragraph{Blame everybody.} Evaluating the exploitability of an HTTP parsing bug is inherently context-sensitive.
    A new origin server could be released tomorrow that makes a previously benign transducer bug exploitable.
    Consequently, some have suggested that no party should be held responsible for exploitable parsing discrepancies~\cite{treqs}; we argue instead that \textbf{all} parties should be held responsible.
    When a standard exists, implementations are obligated to follow it.
    Anything less than strict compliance leaves the door open to the classes of attacks described in this paper.

    \paragraph{Do HTTP/2 and HTTP/3 Solve the Problem?} Potentially.
    Enforcing the use of end-to-end HTTP/2 or HTTP/3 would eliminate many of these vulnerabilities, particularly those caused by the chunked transfer encoding.
    However, web standards last a long time, and the need for HTTP/1.1 compatibility will almost certainly remain for many years to come.
    Further, end-to-end HTTP/2 and HTTP/3 are less susceptible to parser discrepancy-related attacks because there are simply fewer HTTP/2 and HTTP/3 implementations than there are HTTP/1.1 implementations.
    This is presumably because it is straightforward to write an HTTP/1.1 server that works well enough most of the time.

    \paragraph{Future Work.} While the behaviors of transducers are well-studied from the user-facing side, it remains to be seen whether forwarding malformed responses through transducers in the other direction can also be a source of exploitable discrepancies.
    Applying these techniques to other protocols where servers accept and forward messages in chains, such as BGP~\cite{kent2003securing}, may also be possible.
    It may also be worth exploring to what extent removing support for little-used, dangerous constructs, such as chunk extensions, from HTTP implementations would lead to improved security outcomes, despite violating the RFCs.

\section{Acknowledgements}

We'd like to thank the SafeDocs teams at Dartmouth College, Narf Industries, Trail of Bits, and Galois. We'd also like to thank Google and Fastly for their generous bug bounty programs.

\bibliography{refs}
\appendix

\end{document}